**Y. Ishikawa, R. Hattori, Y. Yao, D. Katsube, and K. Sato**

**High-throughput, non-destructive, three-dimensional imaging of GaN threading dislocations with in-plane Burgers vector component via phase-contrast microscopy**

**High-throughput, Non-Destructive, Three-Dimensional Imaging of GaN Threading Dislocations with in-Plane Burgers Vector Component via Phase-Contrast Microscopy**

Yukari Ishiakwa[1,*], Ryo Hattori[2], Yongzhao Yao[1,3], Daiki Katsube[1], Koji Sato[1]
[*]yukari@jfcc.or.jp

[1]Japan Fine Ceramics Center, 2-4-1, Mutsuno, Atsuta, Nagoya, 456-8587, Japan
[2]Ceramic Forum Co., Ltd., 3-19-6, Kandanishiki-cho, Chiyoda, Tokyo 101-0054, Japan
[3]Mie University, 1577, Kurimamachiyacho, Tsu, 514-8507, Japan

We demonstrate a nondestructive, high-throughput method for observing dislocations in GaN (0001) using phase-contrast microscopy (PCM). The PCM images (359 × 300 μm $^2$) analyzed in this study were acquired with an exposure time of 3 ms per image. The one-to-one correspondence between threading dislocation (TD) contrasts in PCM images and the corresponding contrasts in multiphoton excitation photoluminescence (MPPL) images provides clear evidence that PCM can detect TDs with in-plane Burgers vector components. The contrast shape in PCM reflects the inclination of dislocations with respect to the surface normal: dot contrasts correspond to vertical dislocations, whereas line contrasts correspond to inclined dislocations. By shifting the focal plane from the top surface to the back surface, the three-dimensional propagation paths of dislocations can be visualized. The PCM image obtained represents a projection of threading dislocations within a thickness of approximately 43 μm. Dislocations spaced as close as 1.3 μm can be individually resolved. In addition, the capability of PCM to detect scratches, subsurface scratches, facet boundaries, and voids was demonstrated. This study establishes PCM as a versatile and laboratory-accessible technique for three-dimensional, nondestructive characterization of dislocations and other defects in wide-bandgap semiconductors.

## I. INTRODUCTION

Gallium nitride (GaN) is a promising wide-bandgap semiconductor for blue light–emitting devices and next-generation power electronics owing to its favorable physical properties, such as a direct bandgap, wide bandgap energy, high breakdown voltage, and high frequency response[1,2,3,4]. However, dislocations in GaN can severely degrade device performance, yield, and lifetime. In fact, several studies have demonstrated that threading screw dislocations act as leakage paths in PN diodes, as revealed by dislocation analysis at emission spots observed using emission microscopy[5,6,7,8,9]. Therefore, nondestructive methods capable of detecting dislocations across the entire area of a GaN wafer are urgently required. In GaN wafers, there are TDs that extend approximately parallel to the [0001] direction (c-axis) and basal-plane dislocations lying on the (0001) plane (c-plane). TDs can be further classified into three types according to their Burgers vectors: threading edge dislocations (TEDs) (with Burgers vectors lying in the c-plane), threading screw dislocations (TSDs)(with Burgers vectors parallel to the c-axis), and threading mixed dislocations (TMDs) (with Burgers vectors having both in-plane and c-axis components).

Cathodoluminescence (CL) observation had long been the only technique available for detecting dislocations in GaN[10,11]. However, this approach is cumbersome because of its inherent limitations, such as restrictions on sample size and observable area, the requirement for vacuum conditions, and the long acquisition time. More recently, Tanikawa et al. demonstrated that multiphoton photoluminescence (MPPL) is a powerful technique for detecting dislocations, as it enables observation under ambient conditions and allows the three-dimensional (3D) propagation paths of dislocations to be visualized[12,13]. Nevertheless, wafer-scale observation using MPPL is impractical because of the prohibitively long acquisition time; MPPL images must be obtained by scanning with an excitation infrared laser across the observation area. Detection of TDs with in-plane components of the Burgers vector by measuring the shift of the $E_2^H$ phonon peak using Raman scattering has also been reported[14,15,16]. Wafer-scale observation of dislocations remains impractical because of the extremely long measurement times required[17].

Other candidate methods for nondestructive dislocation detection include X-ray topography (XRT) and polarization microscopy (PM). Dislocation detection in GaN crystals with low dislocation densities by XRT has been reported[18,19,20,21,22,23,24]; however, the dislocation densities of most GaN wafers exceed the resolution limit of XRT (~$10^4$ $cm^{-2}$). Similarly, although PM can detect dislocations in GaN[25,], it is unsuitable for characterizing their distribution because its resolution (~$10^4$ $cm^{-2}$) is constrained by the large size of the dislocation contrast (50–100 μm)[26,27]. These facts highlight the need for a nondestructive, high-resolution, wide-field imaging method that enables wafer-scale observation of dislocations in GaN.

Recently, phase-contrast microscopy (PCM) has been applied to the detection of dislocations in 4H-SiC[28,29] and $\beta$-$Ga_2O_3$[30]. Because the dislocation contrast dimension in PCM is only a few micrometers—much smaller than that in PM—PCM appears to be a highly promising method for wafer-scale observation of dislocations in GaN.

In this study, (0001) GaN chips were examined using PCM. By establishing a one-to-one correspondence between contrasts in PCM images and dark spots in MPPL images, we confirmed that PCM can detect TDs with in-plane components of the Burgers vector. Furthermore, by shifting the focal plane from the surface to the back side, we demonstrated that PCM enables three-dimensional visualization of dislocation propagation paths. The comparison with the MPPL images confirmed that the PCM image represents a projection of threading dislocations within a thickness of approximately 43 μm, and that dislocations spaced as close as 1.2 μm can be individually resolved. In addition, we identified other types of defects detectable by PCM besides dislocations.

## II. EXPERIMENTAL DETAILS

Chips cut from a 2-inch hydride vapor phase epitaxy (HVPE)-grown (0001) GaN wafer was used. Both surfaces of each chip were chemically and mechanically polished to minimize light scattering. The dislocation density of the chips was $7\times10^5$ $cm^{-2}$. Phase-contrast microscopy (PCM) was carried out using a Crystalline Tester® CP1 (Ceramic Forum Co., Ltd., Japan) equipped with a 20× objective lens (NA = 0.5) with a phase plate and a condenser lens (NA = 0.78) with a ring aperture. Illumination was provided by a 405 nm LED. Although the LED power was limited to 30%, the exposure time for wide-field imaging (359 × 300 $\mu m^2$) was short as 3 ms. Each chip was mounted on a cover glass placed on the sample holder. PCM images were acquired by scanning the entire chip surface in the XY plane while focusing on the top surface to confirm dislocation detection. Three-dimensional PCM imaging was performed by shifting the focal plane of the objective lens from the top surface to the back surface in 12 μm steps. The focal-plane shift is calculated by multiplying the objective lens displacement by the refractive index of GaN (n=2.4). MPPL observations were performed using an A1 $MP^+$ (Nikon, Japan) equipped with a 50× objective lens (NA = 0.8). Excitation was carried out at 800 nm, and the emission was collected at 370 ± 20 nm. The scan time for an image area of 123 × 123 $\mu m^2$ was 32 s. The Z-direction resolution is estimated to be approximately 3.6 μm, since basal-plane dislocations can be detected within a focal shift of 3.6 μm along the Z axis. Three-dimensional MPPL observations were performed by shifting the focal plane of the objective lens from the top surface to the back surface in 1.2 μm steps.

To form etch pits at the positions where dislocations emerged at the top surface, the chips were etched in molten KOH+NaOH at 450 °C. This etching process reveals TD positions near the subsurface, since the etched thickness in dislocation-free regions is less than 1 μm. For transmission electron microscopy (TEM) observation of dislocations, site-specific samples containing the core of an etch pit were prepared using focused ion beam (FIB) microsampling.

## III. RESULTS AND DISCUSSION

### A. Validation of Threading Dislocation Detection by PCM

Figure 1 shows (a) a PCM image, (d) an MPPL image focused on the top surface, and (c) an etch-pit image of the same area on a (0001) GaN chip with a dislocation density of $7\times10^5$ $cm^{-2}$. Dark dots in Fig. 1(d) and dark hexagons in Fig. 1(c) correspond to TDs. Small irregular bright contrasts arise from redeposited Ga due to laser marking, and white arrows in Fig. 1(d) indicate dust. In Fig. 1(a), several dot contrasts (1–14) correspond to dark spots (1–14) in Fig. 1(d), while line contrasts (I–VIII) in Fig. 1(a) appear near the corresponding dark spots (I–VIII) in Fig. 1(d).

Figures 1(e) and 1(f) show MPPL images obtained by shifting the focal plane 12 μm and 24 μm toward the back surface, respectively. Dark spot I, which aligns with spots 1–7 in Fig. 1(d), shifts toward the lower right with increasing depth [Fig. 1(e) and Fig. 1(f)]. Figure 1(b) presents a projection of the MPPL stack from the surface to a depth of 46 μm. In this projection, spots 1–7 appear as short lines, indicating TDs nearly normal to the surface, whereas spot I forms an extended line, indicating an inclined dislocation. Thus, dot and line contrasts in PCM correspond to vertical and inclined TDs, respectively.

The inclination angle $\theta$ from the surface normal can be determined from the projection length (l) and Z-shift ($z_s$) using

$$\tan\theta = \frac{l}{z_s \times n_{GaN}} \ ,$$

yielding ~8°. The lengths and angles of lines I–VIII in Fig. 1(a) are almost consistent with those in Fig. 1(b). Projection images were generated from MPPL stacks acquired at 2.4-μm intervals and subsequently compared with PCM images. The lengths and distributions of inclined dislocations in the 41–43 μm stacked images closely correspond to those observed in PCM images. Based on this agreement, the dislocation contrasts observed in PCM can be interpreted as a projection of TDs within a thickness range of approximately 43 μm.

Figures 2(a) and 2(c) show PCM images, while Figures 2(b) and 2(d) present the corresponding MPPL images. The distances between TDs indicated by white arrows (1–8) range from 1.3 to 2.3 μm, and each dislocation can be individually resolved in the PCM image. In contrast, the distances between vertical TDs indicated by yellow arrows (A–B) range from 0.6 to 1.1 μm, and in which case the dislocations cannot be resolved individually by PCM and instead appear as a single contrast feature. These observations indicate that PCM with the present optical configuration (405 nm light, 20× objective lens, NA = 0.5, and condenser lens with NA = 0.78) can resolve dislocations spaced as close as 1.3 μm.

## B. Detectable and Undetectable dislocations by PCM

In Fig. 1(c), small, medium, and large pits are observed. Small pits (1–14) correspond to TEDs and appear as dots in Fig. 1(a) and as dots or short lines in Fig. 1(b). Medium pits (I–VIII) are inclined dislocations and are observed as lines in both Fig. 1(a) and Fig. 1(b). Large pits (A and B), indicated by pink arrows, are not visible in Fig. 1(a) but appear as dots in Fig. 1(c). TSDs and TMDs are observed as medium or large hexagonal pits, although the relationship between pit size and dislocation type varies depending on the etchant[31,32]. To definitively identify dislocation types, TEM analysis is required. Figure 3 presents TEM images of dislocations VI and A. Figures 3(a) and 3(b) show images of dislocation VI taken with $\boldsymbol{g} = 0002$ and $\boldsymbol{g} = 11\bar{2}0$, respectively. Because dislocation VI is visible under both conditions, it is identified as a TMD. The dislocation is inclined by approximately 8° from the surface normal, consistent with the inclination estimated from the MPPL projection image in Fig. 1(b). Figures 3(c) and 3(d) show images of dislocation A with $\boldsymbol{g} = 0002$ and $\boldsymbol{g} = 11\bar{2}0$, respectively. Dislocation A is identified as a TSD because it is visible with $\boldsymbol{g} = 0002$ but not with $\boldsymbol{g} = 11\bar{2}0$. That is, the TDs undetectable by phase-contrast microscopy are TSDs. Similarly, Raman scattering mapping[14,16] and polarization microscopy (PM) have been reported to fail in detecting TSDs[26], likely because the absence of shear stress in the XY plane for TSDs hampers their detection[27]. This result suggests that PCM detects dislocations through shear stress in the XY plane. Because TSDs serve as leakage sources in GaN devices, clarifying the detection conditions of PCM is an important subject. This can be addressed by combining MPPL and PCM, extracting TDs detected by MPPL but not by PCM to evaluate the density and spatial distribution of TSDs.

Dislocation A is nearly vertical to the surface, as it appears as a dot in the MPPL projection image in Fig. 1(b). The observed propagation directions of dislocations with different Burgers vectors (TED: vertical to the surface, TSD: vertical to the surface, TMD: inclined from the surface normal) are consistent with the report of Tsukakoshi et al[13]. However, estimation of dislocation type based solely on propagation direction should be approached with caution, as propagation depends strongly on the crystal growth process.

The BPDs indicated by black arrows in Fig. 1(e) are not detected in Fig. 1(a), even though they are within the detection range of PCM. This may be because BPDs intersect the optical axis perpendicularly, producing a phase change corresponding only to their own thickness, whereas PCM acquires phase information integrated over a thickness of approximately 43 μm. Consequently, the resulting phase change may fall below the detection sensitivity of PCM, or alternatively, the difference may arise from variations in the strain field depending on the dislocation line direction.

### C. Three-Dimensional Observation of Threading Dislocations

Figures 4(a) and 4(b) show MPPL and PCM images of the same surface area, respectively. The positions of dislocations (1–15) in Fig. 4(a) correspond one-to-one with those in Fig. 4(b). Figures 4(b)–4(h) present PCM images of the same area with the focal plane shifted from the top surface to the back surface in steps of 60 μm. The supplementary video shows PCM image variations as the focal plane was shifted in 12 μm steps. Dislocations 6, 7, and 11 shift toward the lower right, upper left, and right, respectively, as the focal plane is moved deeper into the crystal.

Because MPPL imaging of dislocations inside the GaN chip was difficult due to low PL intensity, we confirmed a one-to-one correspondence between the MPPL image at the back surface [Fig. 4(i)] and the PCM image [Fig. 4(h)]. Figure 4(i) appears as a horizontal flip because it was obtained from the back surface. The positions of dislocations (2–10, 12–15) in Fig. 4(i) agree well with those in Fig. 4(h). These results demonstrate that PCM enables visualization of the three-dimensional propagation of dislocations.

### D. Other detectable defects

The usefulness of PCM as an inspection apparatus increases with the variety of defects it can detect. Dark straight lines (black arrows) and white lines (white arrows) corresponding to scratches and facet boundaries in Fig. 4(i) are also observed in Fig. 4(h). Notably, an indistinct scratch (yellow arrow) in Fig. 4(i), attributed to a subsurface scratch[33], is visible in Fig. 4(h). In contrast, bright tail-like patterns attributed to carrier concentration inhomogeneity in Fig. 4(i) were not detected in Fig.

4(h). In addition, Plate-like voids with sub-micrometer thicknesses and widths of several to more than ten micrometers, as indicated in the MPPL image of Fig. 5(b), were present in the examined sample. These features were likewise detected by phase-contrast microscopy, as shown in Fig. 5(a).

## IV CONCLUSION

We demonstrated that PCM provides a nondestructive and high-throughput method for detecting TDs in GaN (0001). The one-to-one correspondence with MPPL confirmed that PCM is capable of detecting TEDs and TMDs possessing in-plane components of the Burgers vector, whereas TSDs lacking such components cannot be detected. In PCM images, dot contrasts correspond to vertical dislocations, while line contrasts correspond to inclined dislocations. The dislocation contrasts observed in the PCM images can be interpreted as projection of threading dislocations within a thickness range of approximately 43 μm, and PCM can resolve dislocations spaced as close as 1.3 μm. By shifting the focal plane, PCM enabled visualization of the three-dimensional propagation paths of TDs from the top surface to the back surface. In addition, PCM detected scratches, subsurface scratches, facet boundaries, and voids in GaN wafer. These results establish PCM as a practical and versatile technique for nondestructive, three-dimensional inspection of defects in GaN and other wide-bandgap semiconductors.

## ACKNOWLEDGMENT

This study was partially supported by the METI Monozukuri R&D Support Grant Program for SMEs Grant Number JPJ005698 and the Japan Society for the Promotion of Science KAKENHI (Grant No. 23K04444).

## Author Declarations

### Conflict of Interest

The authors have no conflicts to disclose.

### Ethics Approval

This study did not involve human participants or animal subjects and therefore did not require ethics approval.

### AUTHOR CONTRIBUTIONS

**Yukari Ishiakwa**: Conceptualization (eaual); Data curation (lead); Formal analysis (lead); Funding acquisition (equal); Investigation (equal); Writing – original draft (lead); Writing – review & editing (equal).

**Ryo Hattori**; Conceptualization (eaual); Funding acquisition (equal); Investigation (equal); Methodology(lead), Writing – review & editing (equal).

**Yongzhao Yao**; Investigation (equal); Writing – review & editing (equal).

**Daiki Katsube**; Investigation (equal); Writing – review & editing (equal).

**Koji Sato**; Investigation (equal).

### DATA AVAILABILITY

The data that support the findings of this study are available within the article and supplemental material.

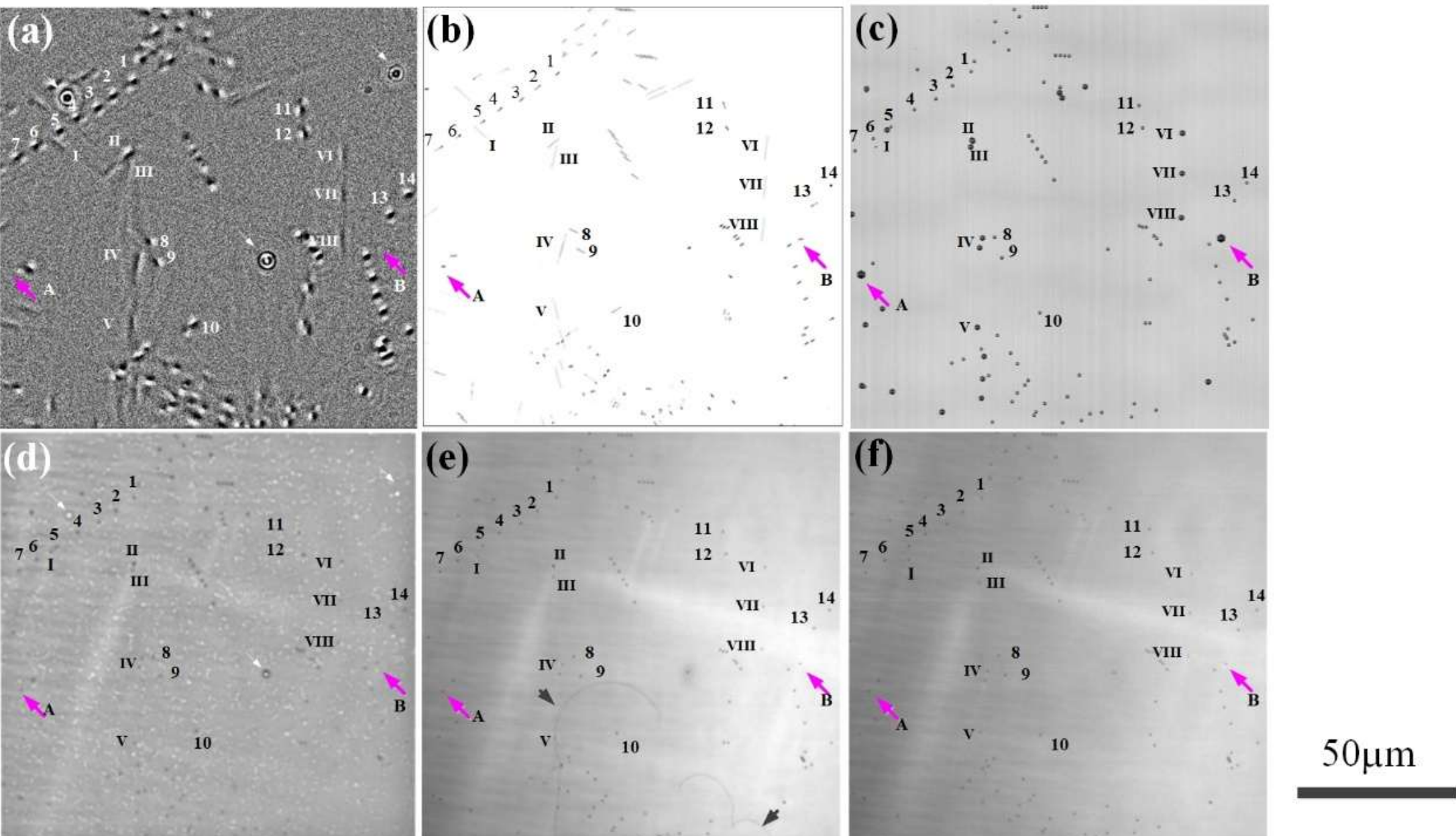

Fig.1 (a) PCM image, (b) projection of MPPL stack from the surface to a depth of 46μm, (c) etch pit image, (d)–(f) MPPL images of the same area. The projection image in (b) was generated by stacking MPPL images acquired from the surface to a depth of 46 μm at 1.2 μm intervals, after dark-spot extraction using the *Find Maxima* function in FIJI[34]. Panels (d)–(f) were obtained at the surface and at depths of 12 μm and 24 μm from the surface, respectively.

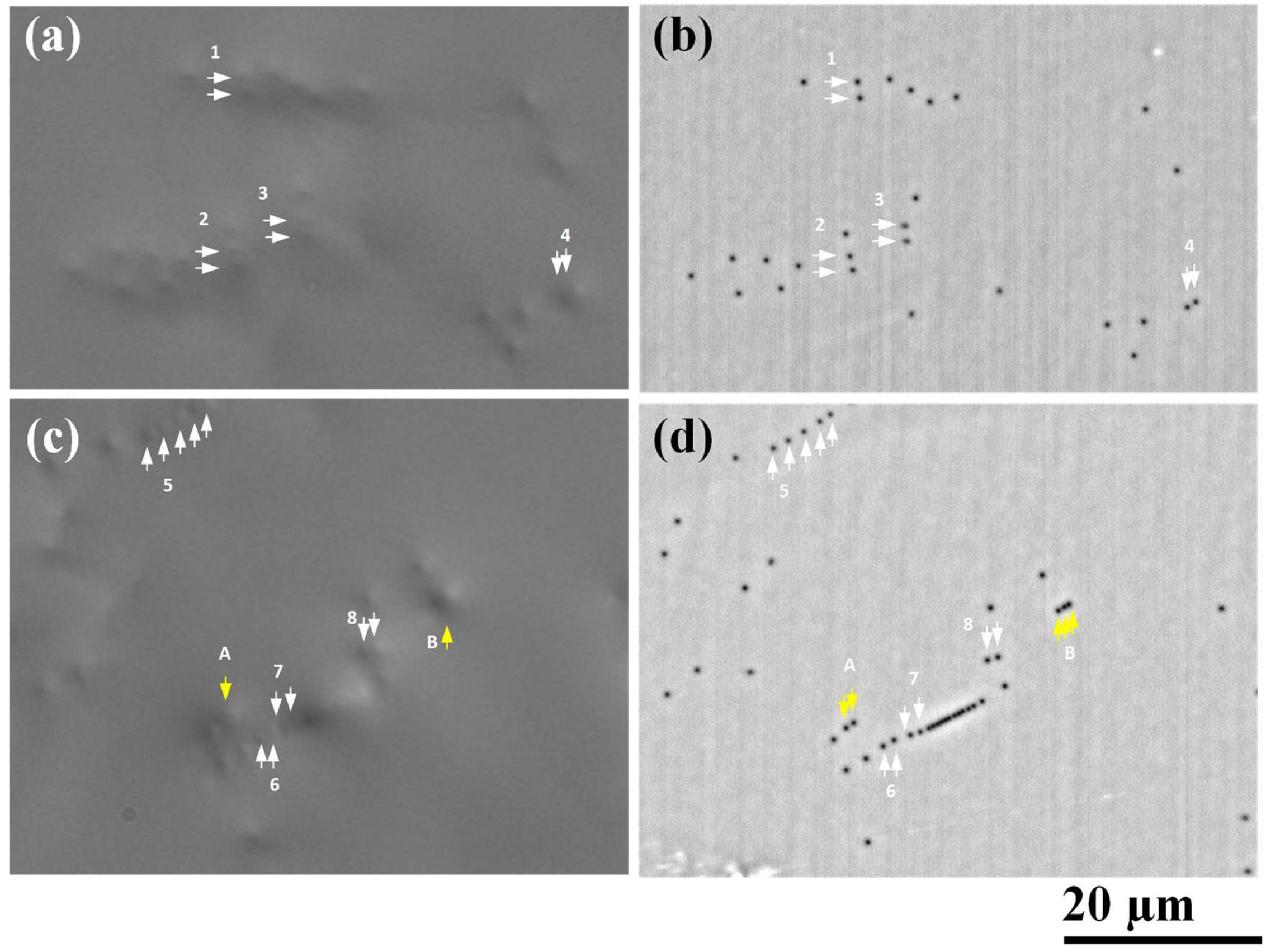

Fig.2 (a) and (c) PCM images, and (b) and (d) the corresponding MPPL images.

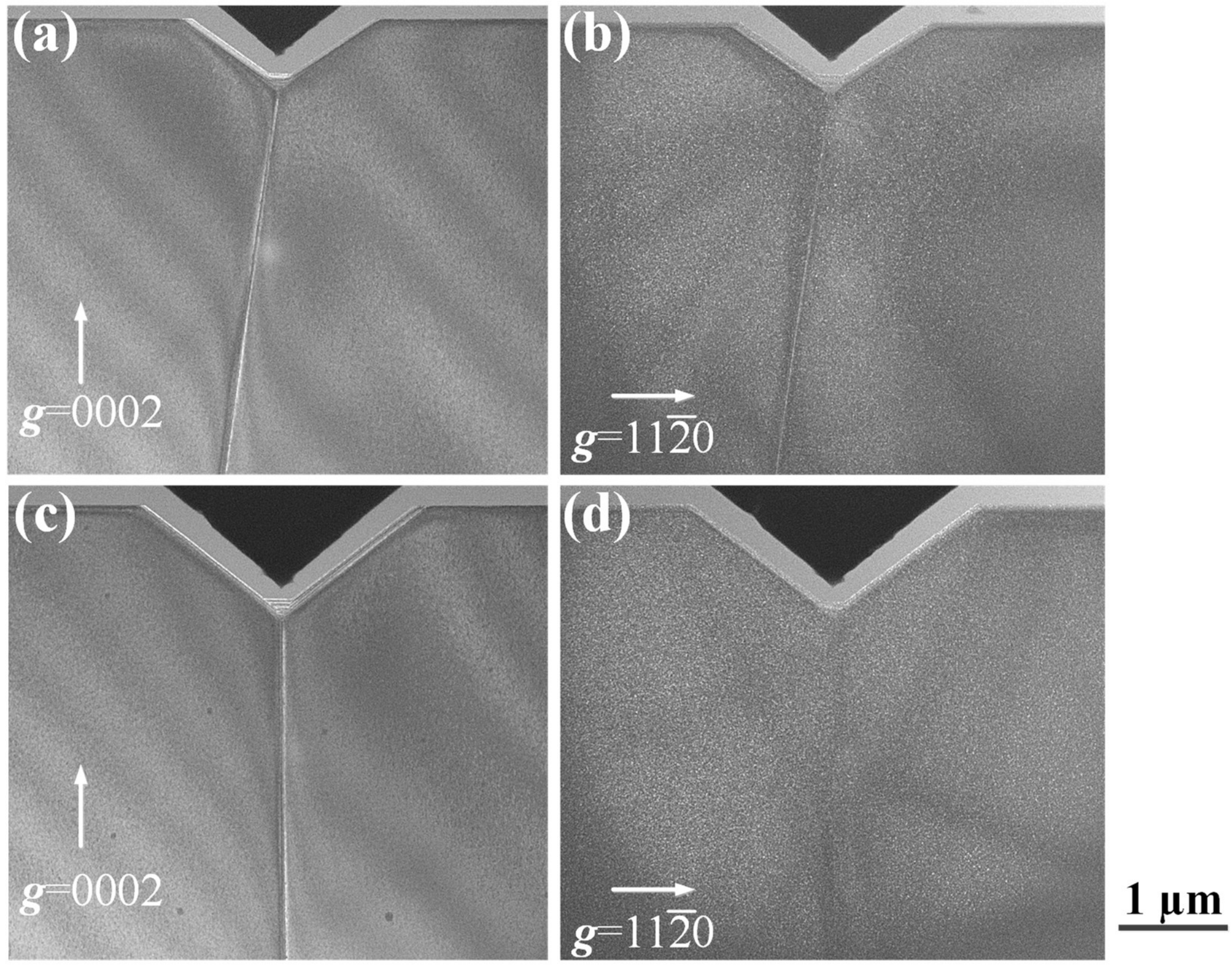

Fig. 3 Cross-sectional TEM images of dislocations VI and A in Fig. 1, shown in (a),(b) and (c),(d), respectively. Panels (a) and (c)were taken with $\boldsymbol{g}$ = 0002, and (b) and (d) with $\boldsymbol{g}$ = 11$\bar{2}$0.

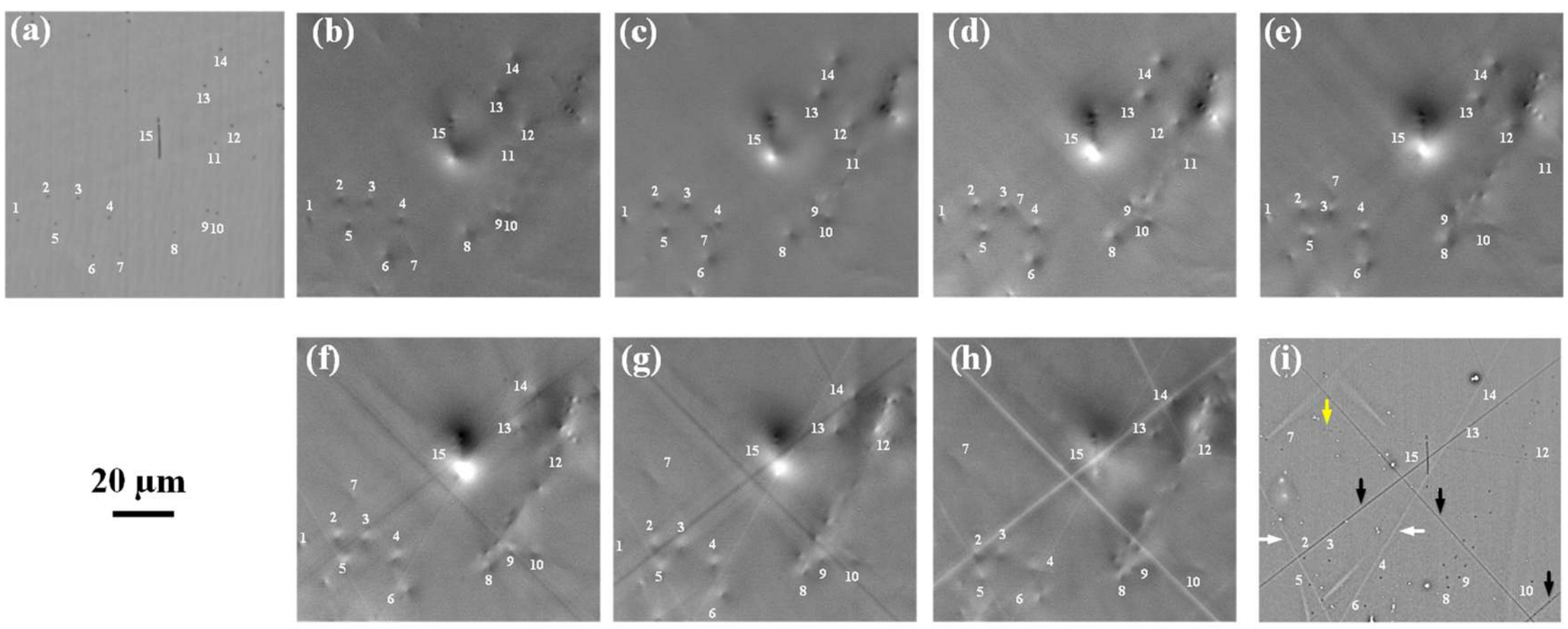

Fig. 4 MPPL images taken at the top surface (a) and back surface (i). Panels (b)–(h) are PCM images obtained from the top surface to the back surface with 60 μm steps.

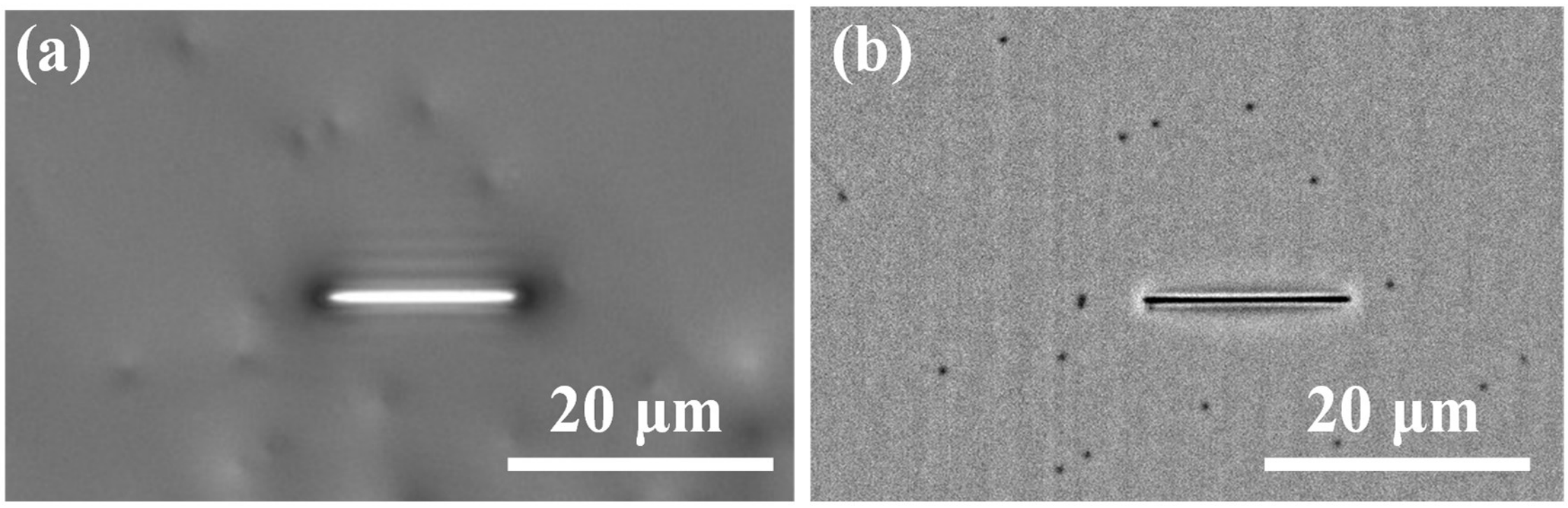

Fig. 5 (a) PCM image and (b) MPPL image of plate-like void.